\documentclass[%
 aip,
 amsmath,amssymb,
 reprint,%
]{revtex4-2}

\usepackage{graphicx}
\usepackage{dcolumn}
\usepackage{bm}
\usepackage{hyperref}
\hypersetup{
    colorlinks=true,
    linkcolor=blue,
    urlcolor  = blue,
    citecolor = blue,
    anchorcolor = blue
    }

\usepackage[utf8]{inputenc}
\usepackage[T1]{fontenc}
\usepackage{mathptmx}
\usepackage{etoolbox}
\usepackage{footmisc}

\makeatletter
\def\@email#1#2{%
 \endgroup
 \patchcmd{\titleblock@produce}
  {\frontmatter@RRAPformat}
  {\frontmatter@RRAPformat{\produce@RRAP{*#1\href{mailto:#2}{#2}}}\frontmatter@RRAPformat}
  {}{}
}
\makeatother

\begin{document}

\preprint{AIP/123-QED}

\title[Spatially resolved stretching-rotation-stretching sequence in flow topology as elementary structure of fluid mixing]{Spatially resolved stretching-rotation-stretching sequence in flow topology as elementary structure of fluid mixing \\}
\author{Ankush G. Kumar}
\author{P. Vishal}
\author{V. Meenakshi}
\author{R. Aravinda Narayanan\textsuperscript{$\dagger$}}
\email[Corresponding author: \textsuperscript{$\dagger$} ]{raghavan@hyderabad.bits-pilani.ac.in}

\affiliation{%
	Department of Physics, Birla Institute of Technology and Science Pilani - Hyderabad Campus, Hyderabad - 500 078, INDIA 
}%

\date{\today}

\begin{abstract}
We performed two-dimensional numerical simulations of passive stirring of two liquids which generated two spatially distinguishable paradigmatic velocity flows, viz. few large extended vortices and rapid oscillations. Using the Okubo-Weiss criterion, we mapped regions of stretching and rotation in the flow field. We find that an elementary oscillatory sequence of stretching-rotation-stretching induces spatial redistribution and steepening of concentration gradients, and the aggregation of such sequences determines fluid mixing at a downstream position. Furthermore, we quantify the paradigms of ‘stretching’ and ‘folding’ by developing measures based on flow topology which would facilitate the design of stirring protocols.
\end{abstract}

\maketitle
This letter concerns the question of elementary structure of mixing in homogeneous fluids, which has been discussed considerably in the literature in the last few decades \cite{ranz1979applications,chate2012mixing, villermaux2003mixing, villermaux2019mixing}. 
When a drop of ink is added to water, the speed at which the state of uniform concentration is reached depends on whether water is stationary or moving; the velocity field determines the rate of mixing. Insights on the mixing process can been gained by either following the velocity field or the passively coupled concentration field \cite{kirby2010micro}.
By picturizing a mixture as discrete interacting sheets of varying concentration profiles shaped by fluid stirring and diffusion, it was shown that mixing is caused by an aggregation of these effects \cite{villermaux2003mixing, villermaux2019mixing}. 

In a fluid stirred through a spatially non-uniform velocity field, idealized fluid elements are shear stretched in one direction (compressed in the other) or rotated, bringing adjacent fluid elements closer to each other, producing steeper concentration gradients to enhance mixing \cite{eckart1948analysis}; in experiments, this is visualized as stretching and folding of material lines \cite{reynolds1894study, ottino1989kinematics, ottino1990mixing}. Recently, fluid deformations delineated through geometric measures showed that mixing occurs in two steps separated in time scales \cite{kelley2011separating}. Initial linear deformations are dominated by stretching which builds up to cause non-linear deformations associated with folding. 

In this Letter, we quantify stretching and folding by developing measures that are directly based on fluid velocity deformations. The advantage of this approach is that mixing arising out of specific stirring protocols can be distinguished. This gains importance, as flow fields, in particular, spatiotemporal chaotic fields \cite{stroock2002chaotic, ober2015active, vagner2019flow, gepner2020use} are being designed through active and passive geometries\cite{bayareh2020active, raza2020review}, to overcome diffusion-limited mixing in low Reynolds number (\textit{Re}) micro-scale flows. In our study, we characterize two prototypical fluid flow topologies using the Okubo-Weiss parameter (Q), which spatially resolves stretching and rotation components in a velocity field \cite{okubo1970horizontal, weiss1991dynamics}. We discover that fluid stirring, and thereby, local mixing, is imprinted in a specific stretching dominated sequential oscillation of Q. This elementary structure of mixing causes resdistribution and sharpening of concentration gradients; the number of such sequences depends on the stirring scheme which aggregates to determine the mixing state at a downstream location. 
\begin{figure*}
\includegraphics[width=\linewidth]{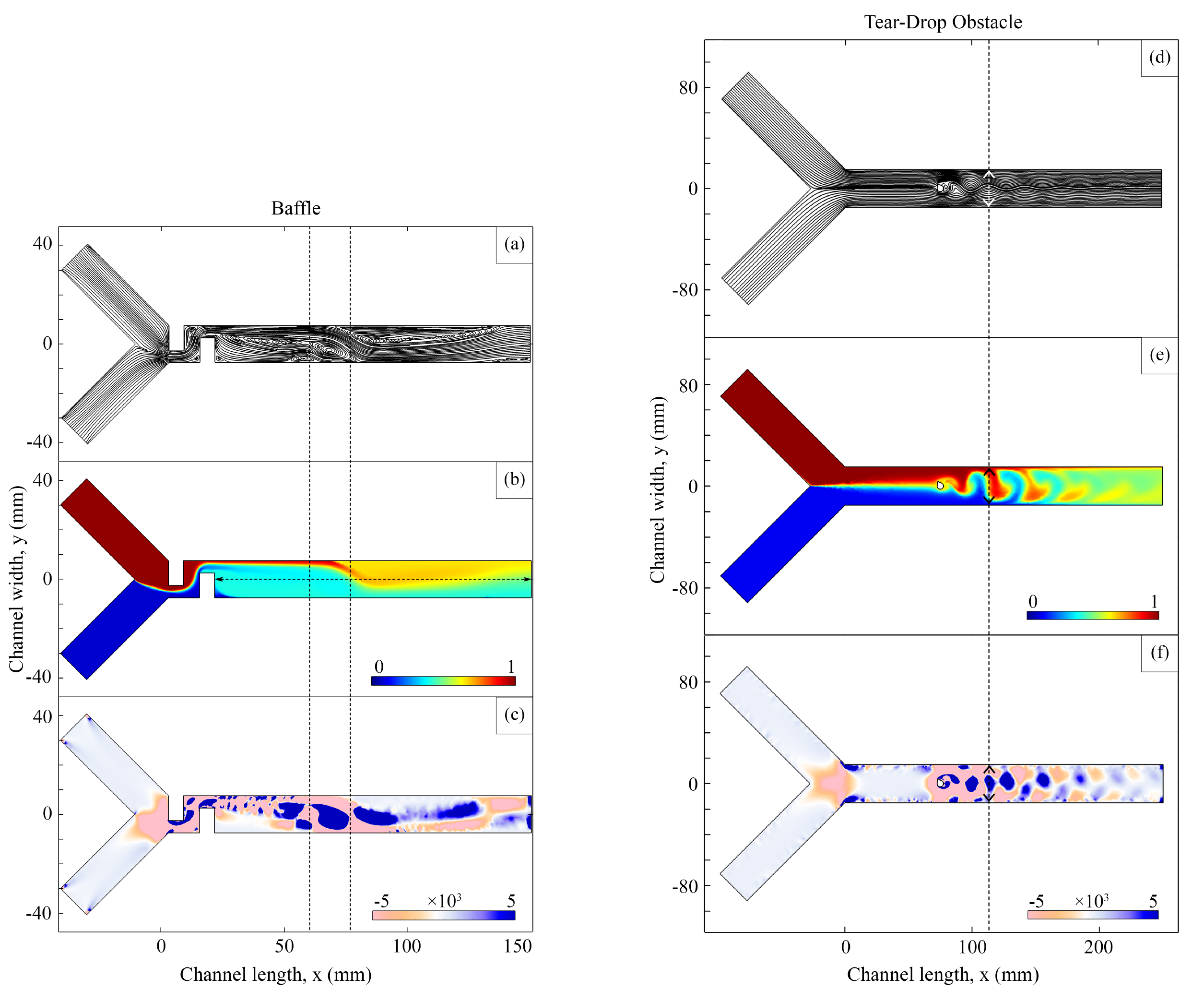}
\caption{\label{fig:1} (color online) \textbf{2D Numerical simulation of passive mixing in BF and TD configuration.} Plotted quantities were extracted at one instance of time after steady state was reached. (a) and (d) show velocity streamlines at \textit{Re} = 90 and \textit{Re} = 37.5, respectively; (b) and (e) show concentration field and in those plots, black regions (brown, and blue online) correspond to unmixed state while light gray (light green online) corresponds to fully mixed state of the two liquids; (c) and (f) show Q map derived from velocity streamlines; black (blue online) represent Q $>$ 0 (rotation) and light gray (pink online) represent Q $<$ 0 (stretching). The dashed lines are discussed in the text. }
\end{figure*}

A close connection exists between the state space of a Hamiltonian system and a two-dimensional incompressible flow field \cite{aref1984stirring, morrison1998hamiltonian}. Hamiltonian dynamics are controlled by two types of critical points, namely, elliptic (centre) and hyperbolic (saddle) \cite{strogatz2018nonlinear}. Ouellette and Gollub showed that by converting flow trajectories of advected particles into curvature fields, the type and spatial location of critical points can be extracted \cite{ouellette2007curvature} using the Q parameter, which is defined as follows:    
\begin{center}
    \begin{equation}\label{equation1}
     Q = \frac{1}{2} (\| \Omega \|^2 - \| S \|^2),
    \end{equation}
\end{center} 
where $\Omega = \frac{1}{2} (\nabla u - \nabla u^T)$ is the vorticity, and $S = \frac{1}{2} (\nabla u + \nabla u^T)$ is the rate-of-strain tensor: The two quantities are the anti-symmetric and symmetric parts of the velocity gradient tensor ($\nabla u$), respectively. They further demonstrated that hyperbolic ($Q_-:Q<0$) and elliptic ($Q_+:Q>0$) critical points corresponding to stretching and rotation dominated regions of the flow, at  low \textit{Re}, forms an ordered square lattice:  As $\textit{Re}$ increases, the lattice gradually loses order, reflecting the deformation of the flow field \cite{ouellette2007curvature}. Thus, it can be gleaned that critical points in state space govern the configuration space flow dynamics \cite{perry1987description}.

Chaos theory helps translate ideas from state space to real flows that have stretched and rotational character: A signature of chaotic mixing is the exponential separation of flow trajectories \cite{ottino1992chaos}. Such flow separation is understood through two types of iterated maps - Bakers maps (iterative stretch and fold) and Twist map (whirling vortices), which help visualize distinct ways to increase the concentration gradient and thereby shorten the diffusion distances \cite{kirby2010micro}. For our study, we chose steady pressure-driven laminar flow for the ease of control of the velocity field, which carries the passive concentration field.The prototypical passive mixing configurations, the pair of baffles (BF) and the teardrop (TD) shaped obstacle\cite{subbarao2020}, simulated in this study, mimic the two flow topologies (Fig.\ref{fig:1}). 

Two-dimensional numerical simulations were performed using COMSOL Multiphysics\textsuperscript{\textregistered}\cite{comsol} software’s microfluidics module.  The governing equations used to model the mixing phenomenon, namely, continuity equation, Navier-Stokes equation, and convection-diffusion equation, were computed recursively with suitable boundary conditions in all the discretized fluid elements in the mixing space. One of the mixing liquids, Rhodamine-B of concentration 1 $mol/m^3$ enters the channel through the left arm of the Y-shaped inlet and water through the other (Fig.\ref{fig:1}). A concentration of  0.5 $mol/m^3$ indicates a completely mixed solution.

\begin{figure*}
\includegraphics[scale=0.5]{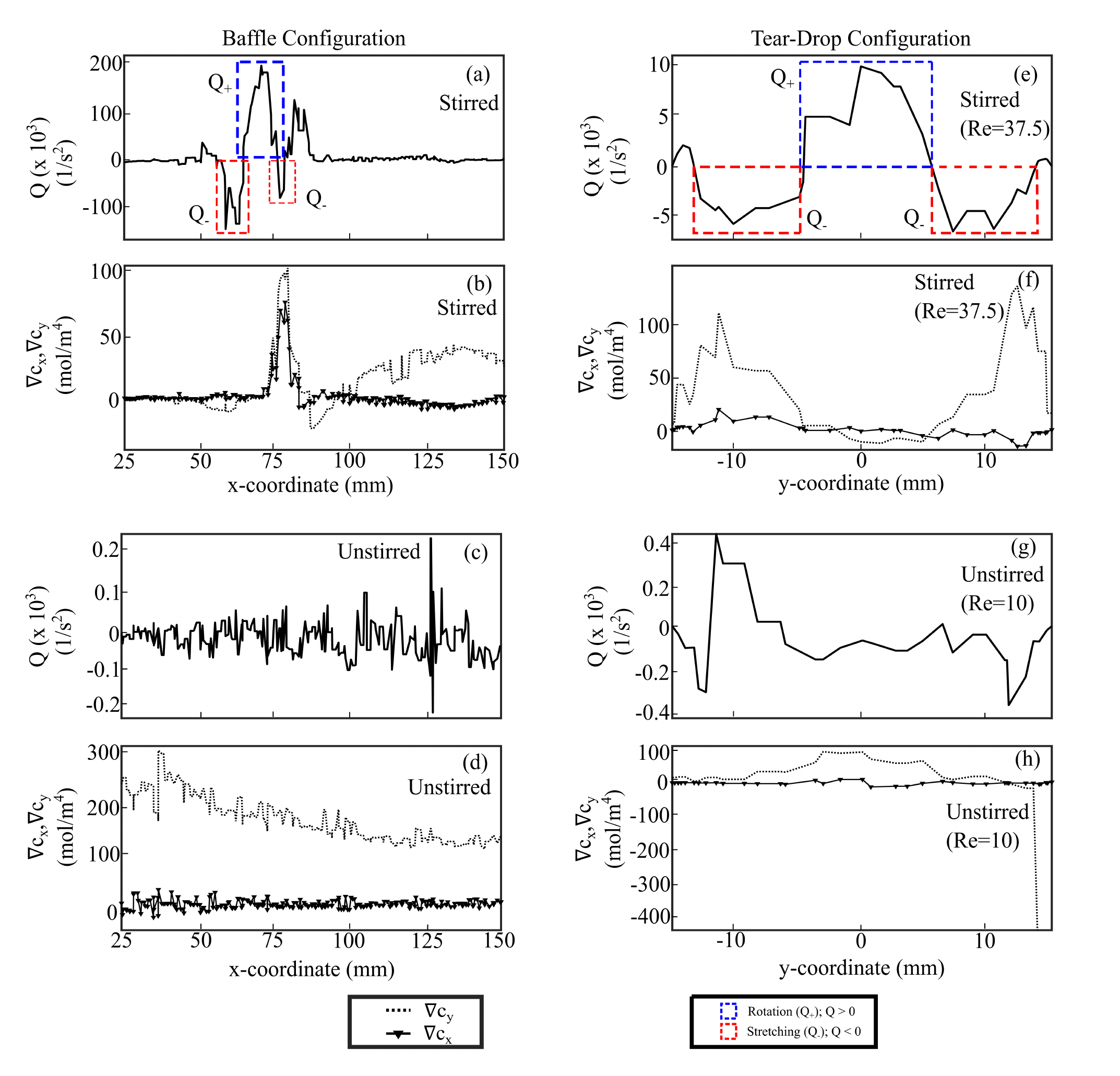}
\caption{\label{fig:2}(color online) Time-averaged spatial dependence of Q for the stirred configurations (a, e), and for the unstirred configurations (c, g). Time-averaged concentration gradient along \textit{x}-axis ($\nabla c_x$) and y-axis ($\nabla c_y$) for the stirred configurations (b, f), and for the unstirred configurations (d, h). The data in the plot for the BF configuration is along \textit{y} = 0 mm (see horizontal dashed line in Fig.\ref{fig:1}(b)) and for the TD configuration along \textit{x} = 118 mm (see the double-arrowed vertical dashed line in Fig.\ref{fig:1}(d)); the choice of the line along which data is presented is decided by the direction of bending of the concentration field. The dashed boxes in (a, e) highlight the $Q_-$$Q_+$$Q_-$-sequence.}
\end{figure*}
\begin{figure*}
\includegraphics[width=\linewidth]{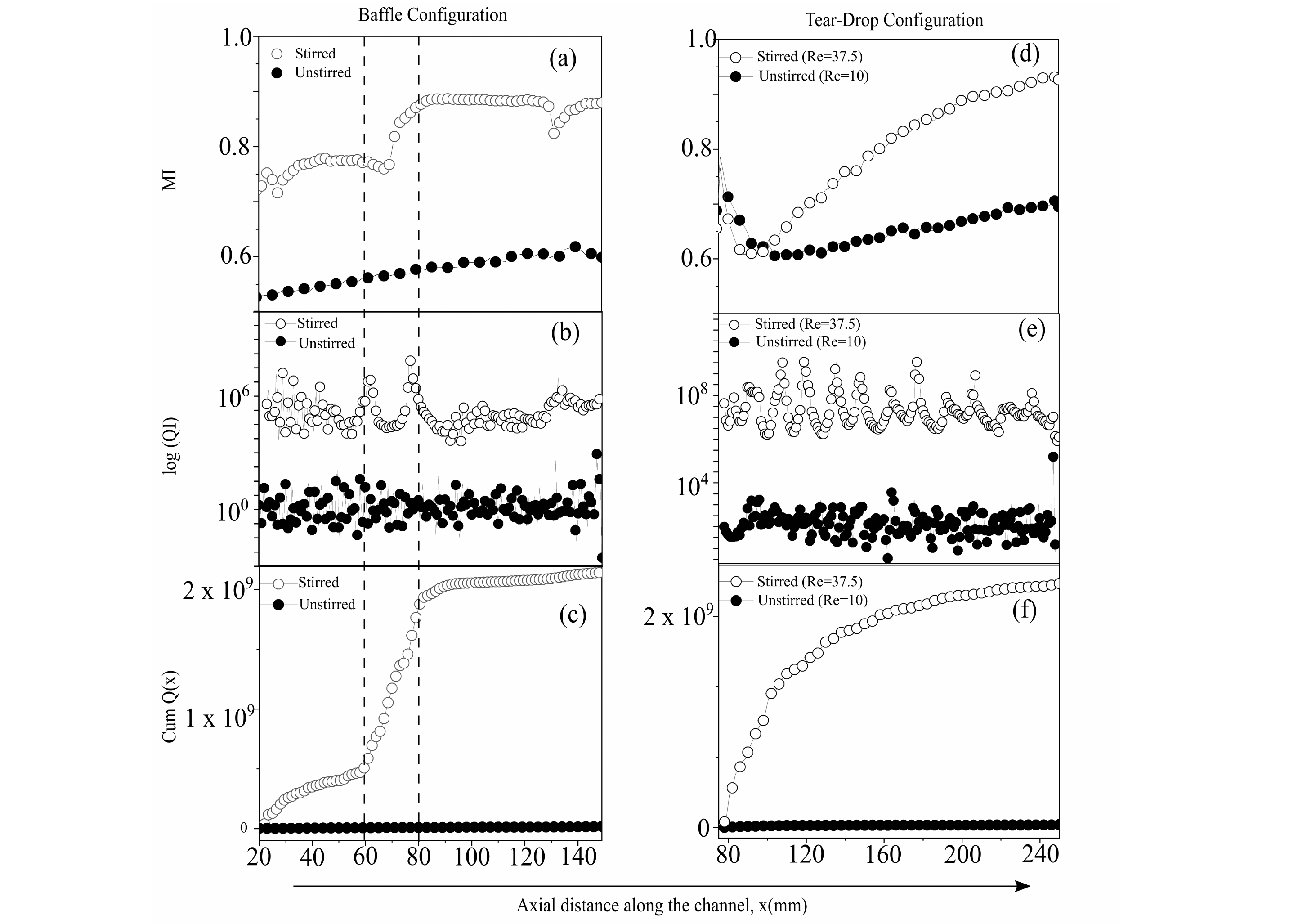}
\caption{\label{fig:3}\textbf{Comparison of mixing index (MI), Q-index (QI) and absolute cumulative Q (Cum Q) for the stirred and unstirred configurations.} In (b) and (e), the unfilled data points have been scaled up to separate the curve from the filled data points. The vertical dashed lines enclose a region having a large vortex (See Fig.\ref{fig:1})}
\end{figure*}

Our focus in this study is on the region beyond the obstacle in the BF and TD configurations (Fig.\ref{fig:1}). While in the BF configuration, we see a few large vortices in the velocity field, contrastingly in the TD configuration the velocity field is oscillatory. As a result, the lines of constant scalar concentration are distorted differently: While in the BF configuration, the concentration field bends about the axial line, in the TD configuration the bending occurs in the transverse direction. The Q-map obtained from Eq.(\ref{equation1}) distinguishes regions of stretching and rotation but at the same time brings to the fore a couple of common features (dashed lines in Fig.\ref{fig:1}). \textit{First}, in the BF configuration, the dashed lines enclose a region having a large vortex and, in the Q-map this corresponds to a sequence of stretching-rotation-stretching ($Q_-$$Q_+$$Q_-$) deformation of the velocity field  (Fig.\ref{fig:1}(c)): In the TD configuration along the dashed line, a similar sequence can be seen (Fig.\ref{fig:1}(f)). The \textit{second} feature is the lag between the distortion of the concentration field and the sequence $Q_-$$Q_+$$Q_-$, which is presented in greater detail in Fig.\ref{fig:2}. 

In state space, the flow topologies shown in Fig.\ref{fig:1} are governed by different sets of critical points. For the BF configuration they are the hyperbolic ($Q_-$) and elliptic ($Q_+$) points. However, if we assume linear damping, $Q_+$ would denote asymptotically stable critical point while $Q_-$ continues to represent the hyperbolic point \cite{strogatz2018nonlinear}. In the TD configuration, initially when \textit{Re} is 10, only asymptotically stable spiral point is present (data not shown): Upon increasing the \textit{Re} to 37.5, the flow stability changes as the system undergoes supercritical Hopf bifurcation giving birth to unstable spiral critical points ($Q_-$) and stable limit cycle ($Q_+$) signifying non-linearity in the flow field \cite{drazin2002introduction}. Here, the limit cycle which is an isolated state space trajectory corresponds to the oscillatory velocity flow in Fig.\ref{fig:1}(d). Subsequently, in this Letter, we call the configurations in Fig.\ref{fig:1} as stirred configurations to compare them with unstirred configurations which are the ‘no baffle’ and the TD at \textit{Re} = 10 configurations.

Concentration gradient causes mixing. In the unstirred configurations, when the two liquids are introduced into the channel, mixing occurs at the interface due to $\nabla c_y$ (Fig.\ref{fig:2}). The corresponding spatial dependence of Q is noisy and has a rate of less than 400/$s^2$. In the stirred configurations, spatially non-uniform velocity field is generated resulting in the oscillating sequence of $Q_-$$Q_+$$Q_-$ which has an amplitude of the order of $10^4$-$10^5$/$s^2$; this causes redistribution, and steepening of concentration gradients downstream. In the stirred BF configuration (Fig.\ref{fig:2}(b)), concentration gradients are sharply localized and are nearly equally distributed among $\nabla c_x$ and $\nabla c_y$, while in the stirred TD drop configuration  (Fig.\ref{fig:2}(f)), redistribution occurs in the transverse direction itself as $\nabla c_y$  ‘splits’ and oscillates along with Q with a distinct phase lag. This resdistribution of concentration gradients upon stirring causes local churning of the fluid, amplifying mixing. We also observed strong correlation between Q and concentration field is observed along axial lines in the BF configuration and along transverse lines in the TD stirred configuration. These results have been confirmed in the time-domain as well. 

To further elucidate the role of the Q oscillations as an elementary structure of fluid mixing, we define three quantities whose variation along the axial line of the flow (x-axis) is plotted in Fig.\ref{fig:3}. The degree of fluid mixing is measured through the Mixing Index (MI):  
\begin{center}
    \begin{equation}\label{equation2}
       MI=1-\sigma
    \end{equation}
\end{center}
Here, $\sigma$ is the standard deviation of the concentration: MI of 0.5 and 1 denotes the unmixed state and mixed states, respectively. Asymmetry in the Q oscillatory sequence suggests that a sum of Q’s in that sequence will be dominated by the stretching component. Therefore, to bring out the relative strength of stretching and rotation of the velocity field, Q-index (QI) is defined:
\begin{center}
   \begin{equation}\label{equation3}
       {QI} = \left( \frac{\Sigma_y Q_-}{\Sigma_y Q_+} \right)^2
    \end{equation} 
\end{center}
Here, $\Sigma_y Q_-$ and $\Sigma_y Q_+$ denote sum over a vertical line in the flow channel. The ‘extra’ stretching deformation accrued from each one of the spatially resolved Q oscillations is captured by the absolute cumulative sum of Q defined as follows:
\begin{center}
    \begin{equation}\label{equation4}
        Cum   Q(x)=\Sigma^{x}_0|\Sigma_y Q_- + \Sigma_y Q_+|
    \end{equation}
\end{center}

Cum Q(x) is computed at every \textit{x}, with an interval of 2 mm, between the inlet and outlet of the channel. In the case of BF configuration, peaks in QI are seen at \textit{x} (mm) = 60, 80, and 130, which result in concomitant discontinuous increases in MI  (Fig.\ref{fig:3}(a)). However, for the TD configuration, since the peaks in QI are many, the closely spaced discontinuous jumps in MI appear as continuous increase of MI  (Fig.\ref{fig:3}(d)). In earlier reports of chaotic mixing due to a spatially varying oscillating velocity field, similar smooth increase in mixing have been seen \cite{stroock2002chaotic, vagner2019flow}. In the unstirred configurations, in which Q oscillation is absent, peaks in QI are conspicuously absent and therefore, MI for them increase linearly with a smaller slope.

We now consider the notion of folding that is associated with non-linear deformations of the velocity field \cite{kelley2011separating}, using Cum Q(x) (Fig.\ref{fig:3}(c), (f)): For the stirred configurations, this quantity closely follows the variations in MI. The explanation is that fluid mixing at an axial location is determined by the state of fluid deformation at preceding upstream positions; they form a recurrence relation in the configuration space. For the unstirred configurations, Cum Q(x) is nearly zero implying presence of \textit {only} linear deformations that cause MI to increase linearly. These results confirm previous findings that fluid mixing due to stirring is an aggregation of stretching events \cite{villermaux2003mixing}.

The similarity in the variation of MI and Cum Q(x) highlights the fact that the mixing process is irreversible because of its dependence on the history of deformations. This irreversibility caused by molecular diffusion destroys spatial correlations among advected particles. As an illustrative example of unstirred flow, consider, experimental demonstration of reversible mixing in Taylor-Couette flow, where a gentle discrete number of clockwise rotations of a highly viscous fluid, that has a large diffusion length, mixes the fluid, and an equal number of anti-clockwise rotation, segregates it \cite{fonda2017unmixing}. The rotation of the fluid, which is a linear deformation, does not shorten the diffusion length and for this process Cum Q(x) will be zero.

In summary, we characterized the complex flow topology that emerges due to stirring, as fluids in an initially segregated state, reach a final, completely, mixed state. We discover that a spatial oscillatory pattern of stretching-rotation-stretching forms the basic building block of mixing; and in effect, a cumulative addition of these building blocks decides mixing at a downstream position. This approach enables us to quantify the century old paradigms of ‘stretching’ and ‘folding’. Furthermore, we establish a link between fluid mixing and the state space of the dynamical system. In the stirred state, flow fields are governed by a pair of critical points corresponding to stretching and rotation dominated regions in the flow. A flow field which has large vortices, described by hyperbolic and elliptic critical points, produces discontinuous increase in mixing and contrastingly, an oscillating field consisting of unstable spiral point and a limit cycle generates monotonic increase in mixing. This work raises the possibility of manipulating these special points to design velocity flow fields to tailor spatial mixing variations\cite{stroock2002chaotic, gepner2020use}.  

\begin{acknowledgments}
Computational resources available at the BITS (Pilani), Hyderabad campus were used for this work.
\end{acknowledgments}


\section*{AUTHOR DECLARATAIONS}
The authors have no conflicts of interest to declare.
\nocite{*}
\bibliography{ankush-bib}

\end{document}